# Developing a seismic pattern interpretation network (SpiNet) for automated seismic interpretation

Haibin Di[1,2]

[1]Georgia Institute of Technology, School of Electrical and Computer Engineering, Center for Energy and Geo Processing (CeGP) at Georgia Tech and KFUPM, Atlanta, Georgia 30308, USA.
[2]Schlumberger, Houston, Texas 77056, USA
E-mail: hdi7@gatech.edu

## ABSTRACT

Seismic interpretation is now serving as a fundamental tool for depicting subsurface geology and assisting activities in various domains, such as environmental engineering and petroleum exploration. In the past decades, a number of computer-aided tools have been developed for accelerating the interpretation process and improving the interpretation accuracy. However, most of the existing interpretation techniques are designed for interpreting a certain seismic pattern (e.g., faults and salt domes) in a given seismic dataset at one time; correspondingly, the rest patterns would be ignored. Interpreting all the important seismic patterns becomes feasible with the aid of multiple classification techniques. When implementing them into the seismic domain, however, the major drawback is the low efficiency particularly for a large dataset, since the classification need to be repeated at every seismic sample. To resolve such limitation, this study first present a seismic pattern interpretation dataset (SpiDat), which tentatively categorizes 12 commonly-observed seismic patterns based on their signal intensity and lateral geometry, including these of important geologic implications such as faults, salt domes, gas chimneys, and depositional sequences. Then we propose a seismic pattern interpretation network (SpiNet) based on the state-of-the-art deconvolutional neural network, which is capable of automatically recognizing and annotating the 12 defined seismic patterns in real time. The impacts of the proposed SpiNet come in two folds. First, applying the SpiNet to a seismic cube allows interpreters to quickly identify the important seismic patterns as input to advanced interpretation and modeling. Second, the SpiNet paves the foundation for deriving more task-oriented seismic interpretation networks, such as fault detection. It is concluded that the proposed SpiNet holds great potentials for assisting the major seismic interpretation challenges and advancing it further towards cognitive seismic data analysis. Meanwhile, more work is expected for evolving the SpiNet by integrating transfer learning, defining more seismic patterns, feeding more seismic datasets and training labels, and optimizing the network architectures.

## INTRODUCTION

As a fundamental tool for characterizing subsurface geology, three-dimensional (3D) seismic interpretation plays a crucial role in various disciplines, such as civil engineering, geohazard assessment, and energy exploration. Interpreting a seismic volume is a time-consuming and labor-intensive process and often requires mutual collaborations between geologists, geophysicists, petrophysicists, and more. Manual interpretation has been the most straightforward and effective approach for solving this problem, in which an interpreter visually analyzes the seismic reflection patterns, identifies the important patterns, and labels them by distinct marks and/or colors. However, the dramatically increasing size of 3D seismic surveying is now significantly challenging the efficiency of such manual interpretation.

For accelerating the interpretation process, geoscientists have made great efforts into developing a full suite of computer-aided tools, such as edge detection, geometry estimation, facies analysis, object extraction, and more. However, most of these tools are designed for interpreting one or some certain features by analyzing seismic signals from different perspectives. Correspondingly, the rest features present in a seismic dataset would be undesirably ignored. For example, as the first edge-detection tool, the coherence attribute (Bahorich and Farmer, 1995) estimates the lateral similarity of seismic waveforms and thereby is effective in depicting the faults and stratigraphic features that obviously break the waveform continuity. Since its popularity, a number of variations and schemes have been developed in improving such attribute (e.g., Luo et al., 1996; Marfurt et al., 1998; Gersztenkorn and Marfurt, 1999; Cohen and Coifman, 2002; Tingdahl and de Rooij, 2005; Di and Gao, 2014; Wang et al., 2016). While clearly highlighting the major faults of apparent displacements, however, most of the edge-detection tools are less efficient for subtle structure interpretation, such as fracture characterization and facies analysis, in which the lateral variation of seismic signals is subtle and beyond the resolution of edge detectors. Detailed summaries of the edge detection can be found in Chopra (2002), Kington (2015), and Di and Gao (2017a). For the purpose of detecting the small-scale structures like subtle faults and fractures, geophysicists then turn to evaluating the variation of the geometry of seismic reflectors, which successfully link the fractures with the high-order reflector geometric attributes, such as curvature (Roberts, 2001) and flexure (Gao, 2013). A suite of schemes is also available for such geometry estimation, whose efficiency in identifying planar seismic structures, such as fractures, has been documented in various case studies (e.g., Di and Gao, 2014b, 2017b; Gao and Di, 2015; Yu and Li, 2017a, 2017b; Qi and Marfurt, 2017). However, such geometric analysis often fails for stratigraphic features, such as channels, reefs, lobes, and overbanks. Instead, accurate stratigraphic interpretation becomes possible by performing seismic facies analysis, particularly the GLCM analysis that estimates the local arrangement of seismic amplitudes in 3D space (Gao, 1999; Eichkitz et al., 2013; Di and Gao, 2017c). The GLCM tool is based on the fact that rock particles are packed in different ways with the depositional environment varying, and correspondingly, the reflection patterns are locally different in terms of their amplitude, frequency, and/or phase.

While depicting the target seismic pattern from the surrounding ones, however, these techniques fail to extract themas separate objects that can be readily fed into framework construction and modeling. For example, a salt body can be visually depicted as high homogeneity and low contrast in GLCM maps (Gao, 2003), but isolating it from the surrounding patterns requires additional tools used in computer graphic and imaging processing. For example, normalized cuts (Lomask et al., 2007) detects salt domes by solving a global optimization problem. The active-contour-models method (Shafiq et al., 2015) starts with the initial boundary from interpreters and then gradually deform it to fit the salt boundary observed in the attribute image. Wu (2016) incorporates discrete pickings by an interpreter into the detection process to guide accurate delineation of salt boundaries, especially in complicated zones with gaps or outliers. Ramirez et al. (2016) adopt the theory of sparse representation and apply it to automatically segment salt structures from 3D seismic dataset. Similarly, (semi-)automatic fault extraction has been popular in the past years with numbers of algorithms presented in this field, including ant tracking (Pedersen et al., 2002), Hough transform (AlBinhassan and Marfurt, 2003), eigenvector analysis (Barnes, 2006), dynamic time wrapping (Hale, 2013), motion vector (Wang et al., 2014), and more.

However, such object extraction often works for one certain structure at a time. Therefore, for a seismic dataset of multiple important structures such as faults and salt-bodies, both algorithms have to be performed individually, which doubles the required interpretation time and efforts. With the success of machine learning in audio/image/video understanding, various labeling and classification techniques have been introduced into the field of seismic interpretation, including facies analysis (Zhao et al., 2015), salt-body delineation (Alaudah et al, 2017; Di and AlRegib, 2017; Di et al., 2018a), and fault detection (Zheng et al., 2014; Huang et al., 2017; Di et al., 2017, 2018b; Alaudah and AlRegib, 2017). A comprehensive overview of machine learning in seismic interpretation can be found in AlRegib et al. (2018). Although the previous studies focus on certain seismic patterns, such classification algorithm can be easily extended for full-pattern interpretation, given a training dataset of all the important patterns well interpreted and annotated.

As observed in our testing, it is very efficient for training an optimal classifier from 3D seismic data; however, applying it for a seismic volume is a time-consuming process, particularly for large datasets, since the classification has to be repeated at every sample in the volume. To resolve this challenge, this study proposes developing a seismic pattern interpretation network (SpiNet) that is capable of understanding and annotating the important patterns in a seismic dataset in real time. The paper is structured as follows. First, we manually annotate the available open-source seismic datasets and present a seismic pattern interpretation dataset (SpiDat) that tentatively categorizes 12 commonly-observed seismic patterns, including these of great geologic importance such as faults and salt domes. Second, we illustrate the SpiNet architecture and describe the training process in detail. Finally, we verify the added values of the proposed SpiNet through two applications: real-time annotation of the F3 seismic volume over the Netherlands North Sea and fault detection from the Great South Basin (GSB) dataset in New Zealand. Based on the results, we draw conclusions at the end of the paper.

## SEISMIC PATTERN INTERPRETATION DATASET (SPIDAT)

The primary goal of seismic interpretation is to understand seismic signals, categorize them into various patterns, connect each pattern with a specific depositional event, and finally reconstruct the geologic history. Therefore, the emerging machine learning techniques, particularly the convolutional neural networks, appear most suitable for tackling the problem of annotating various patterns existing in a seismic dataset. However, we identify three major challenges while building such a neural network. The first challenge comes from the lack of open-source seismic datasets, which are comprehensive and representative enough for all types of seismic patterns. Secondly, most of these available datasets are without interpretational annotations and thus cannot be readily used for supervised network training, validating, and testing. Seismic pattern annotation is precisely the third and largest challenge, which results from the difference between the natural images and the seismic signals. The former typical records the already-known objects such as animals and cars, and thus it is relatively easy for describing these objects and defining a list of them. The seismic signals, however, reflect the subsurface geology that not only requires multiple descriptors (e.g., reflection intensity, lateral geometry, and geomorphology as listed in Table 1) in describing them, but also is complicated in nature and remains unpredictable for seismic interpreters particularly in the frontier exploration areas where the existing seismic surveying technology fails to accurately image the geology. Moreover, the varying interpretational goals further add the difficulty of cohering the seismic pattern description, since a seismic pattern may be named in different ways for different study area and/or with the interpretation focus shifting from one to another even for the same area.

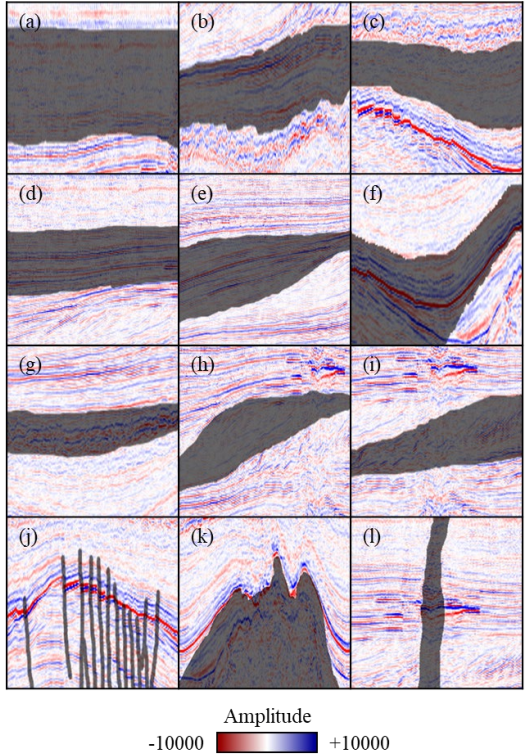

**Figure 1.** Examples of the tentatively-defined 12 seismic patterns, including 7 types of horizons (a-g), 2 stratigraphic sequences (h-i), and 3 structures (j-l). Specifically, the horizons are characterized by two descriptors: intensity (weak or strong) and geometry (flat, dipping, or deformed). Two sequences are regression and transgression, while the three structures are fault, saltbody, and gas chimney.

Considering the essence of manual annotations for implementing machine learning to the domain of seismic interpretation, in this study we propose building a comprehensive seismic pattern interpretation dataset, here denoted as SpiDat, in which the most common and important seismic patterns are defined and annotated. Apparently, it is a long-term project and requires collaborations between geologists, geophysicists, and seismic interpreters. Here, we initialize the SpiDat by utilizing the open-source seismic datasets available for us, including the F3 block over the Netherlands North Sea, the Teapot dome in Wyoming in USA, and the Great South Basin in New Zealand. A total of 12 seismic patterns are tentatively categorized from them, including

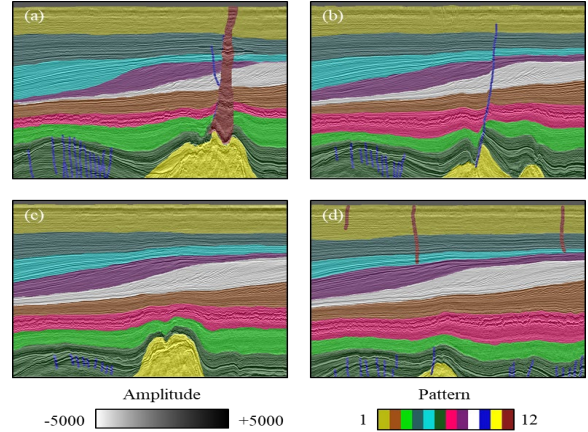

**Figure 2.** Four annotated sections in the seismic pattern interpretation dataset (SpiDat), including inline #190 (a), #290 (b), #390 (c), and #r490 (d) from the F3 block over the Netherlands North Srea. The manual annotation is overlaid over the original seismic amplitude (in gray scale) for the convenience of visualization here.

7 types of horizons, 2 types of stratigraphic sequences, and 3 types of structures (Figure 1). Figure 2 displays the annotations of 4 inline sections (#190, #290, #390, and #490) from the F3 dataset, which is considered most comprehensive among all the available seismic datasets.

## SEISMIC PATTERN INTERPRETATION NETWORK (SPINET)

For efficient subsurface interpretation from large 3D seismic datasets, this study proposes developing an end-to-end seismic pattern interpretation network, here denoted as SpiNet, which is superior over the existing workflows/algorithms in three aspects. First, it is not specifically designed for certain seismic pattern, but maximumly mimics the vision and intelligence of seismic interpreters and thus is capable of identifying as many patterns as an interpreter specifies. Second, it inherits the efficiency of the state-of-the-art deconvolutional neural network (Noh et al., 2015) and thus is capable of annotating a seismic dataset in real time. Third, it is applicable to any seismic dataset of any dimensions. In the section below, we describe the SpiNet in detail, including its architecture, training process, and performance analysis.

### Architecture

Figure 3 illustrates the architecture of the proposed SpiNet in U-shape. It starts with 3 convolutional blocks, which contain 3, 2, and 1 convolutional layers, respectively. Correspondingly, 3 deconvolutional blocks are placed in the end for size recovery to ensure that the annotations are provided at the correct locations in the output image. In the middle is a 1x1 block of 2 layers and 1024 features for connecting the convolutional and deconvolutional blocks. Given a seismic section of size ($M \times N$), at the stage of convolution, with the 1-channel input image going through the 3 convolutional blocks, a series of 2x2 convolution masks extract 32, 64, and 128 features, and meanwhile maximum pooling gradually reduce the size to ($\frac{M}{2} \times \frac{N}{2}$), ($\frac{M}{4} \times \frac{N}{4}$), and ($\frac{M}{8} \times \frac{N}{8}$), respectively. When turning to the stage of deconvolution, with the 1024 features of size ($\frac{M}{8} \times \frac{N}{8}$) going through the 3 deconvolutional blocks, a series of 2x2 transposed convolution merges them into 128, 64, 12 features and meanwhile gradually increases their size to ($\frac{M}{4} \times \frac{N}{4}$), ($\frac{M}{2} \times \frac{N}{2}$), and ($M \times N$), respectively. Meanwhile, in every deconvolutional layer, the corresponding convolutional features are added to the generated deconvolutional features after the transposed convolution. For example, the 128 convolutional features generated from the 3rd convolutional block is used for providing the 128 deconvolutional features from the 1st deconvolutional block. Note that the 3rd and last deconvolutional block produces 12 features, each of which corresponds to one of the 12 seismic patterns defined in Figure 1.

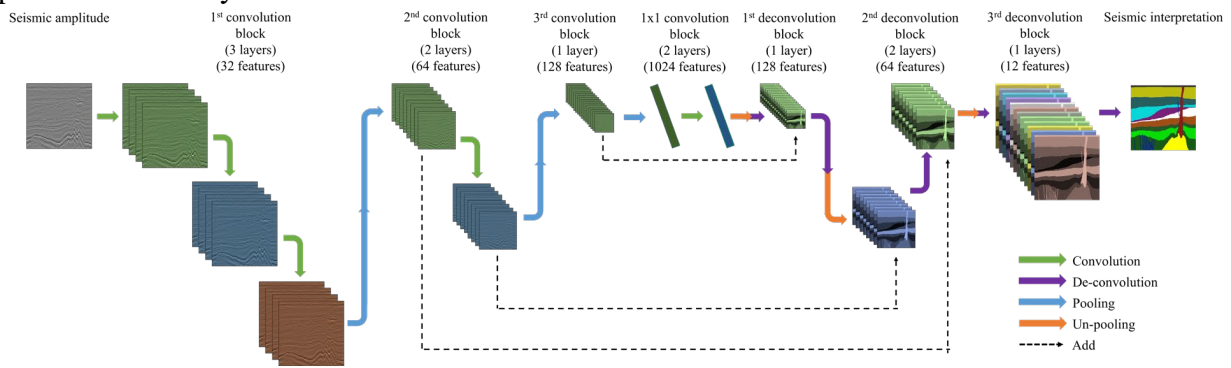

**Figure 3.** The architecture of the proposed seismic pattern interpretation network (SpiNet), which consists of 3 conventional blocks and 3 corresponding deconvolutional blocks.

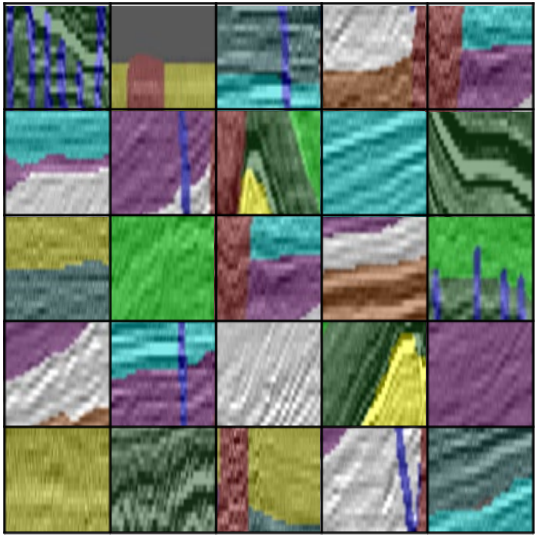

**Figure 4.** 25 examples of the 76,950 training images used for the SpiNet training.

**Training process**

Based on the annotated dataset SpiDat, we then train the proposed SpiNet using 3 of the 4 inline sections in Figure 2, which are #190, #290, #490, whereas inline #390 is reserved for testing the performance of the trained network in the next section.

Considering the insufficiency of manual annotations, this study does not direct feed the 3 sections into the SpiNet training but applies three data augmentation approaches to increasing the amount of training images. First, each inline section is split into small patches of 51 crosslines by 51 samples. 25 example patches are displayed in Figure 4. Second, each small patch is shifted both laterally by 10 crosslines and vertically by 10 samples, which increases the patch amount by 25 times. Third, we rotate these patches as well as their annotation in 5 ways, including 90º-, 180º-, 270º-rotation, and up-right and left-right flipping. Such data augmentation helps provide a total of 76,950 training images from the 3 annotated training sections.

The training of the SpiNet model aims at minimizing the difference between the prediction $P$ and the manual annotation $A$ of all the prepared training patches. There exists a suite of mathematic approaches for quantifying the difference, and the traditional multi-class cross-entropy loss function is adopted in this study (Equation 1).

$$L = -\frac{1}{N}\sum_{n=1}^{N}\sum_{k=1}^{K} A_n(k,O) \cdot \log P_n(k) \quad (1)$$

in which $N$ and $K$ denote the numbers of image pixel and target patterns, respectively. At a given pixel $n$, $A_n(k,O)$ represents the binary indicator (0 or 1) if class $k$ is the correct annotation $O$, whereas $P_n(k)$ represents the probability of the predicted annotation being class $k$. Specifically, $N = 2601$ and $K = 12$ in this study. The typical Adam optimizer is used for loss minimization.

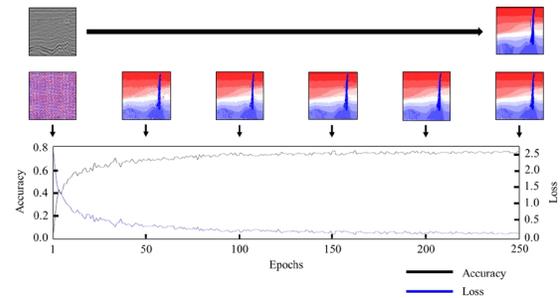

**Figure 5.** The learning curve of training the SpiNet in 250 epochs. It starts from random noises and gradually becomes capable of identifying the outlines of the seismic patterns at epoch #50. The following is to fine-tune the prediction in local with the accuracy increasing to 0.8 and the loss reducing to 0.2 when the training stops after 250 epochs. The performance is expected to improve further by more training epochs.

**Performance analysis**

Figure 5 illustrates the training process in 250 epochs. It starts from random noises, and thus the initial predictions are of wrong patterns. With the training continuing, after about 50 epochs the SpiNet becomes capable of depicting the outlines of the target seismic patterns. The following training is less obvious but gradually improves the accuracy, reduces the loss, and refines the annotations in local. Finally, an accuracy of 0.8 is achieved with the loss decreased to $< 0.2$

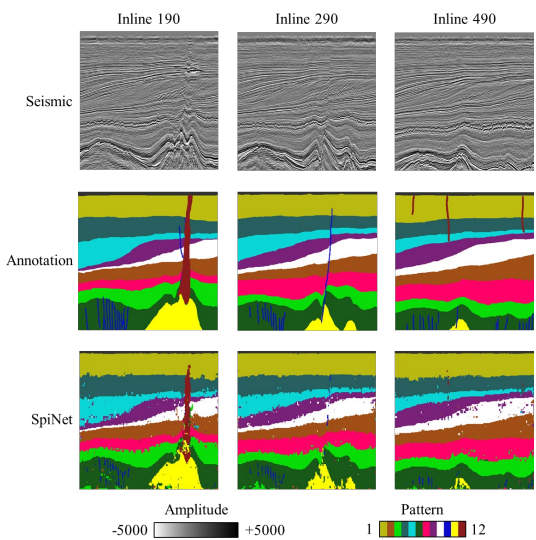

**Figure 6.** The comparison of the manual annotation and the SpiNet prediction on 3 training sections, including inline #190, #290, and #490 of the F3 dataset. Note the good delineation of the important seismic patterns such as the two stratigraphic sequences (in purple and white), the chaotic reflection (in magenta), the salt dome (in yellow) as well as the overlying strong and deformed reflection (in green). More training is expected for further improving the prediction particularly on the faults (in blue) and gas chimney (in red) in sections #290 and #490 that are not well delineated.

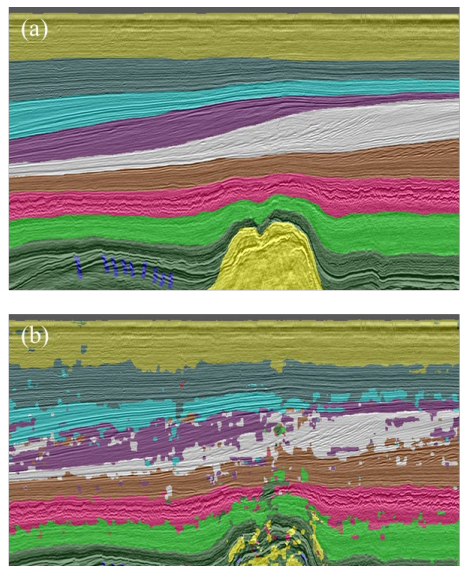

**Figure 7.** The comparison of the manual annotation and the SpiNet prediction on 1 testing sections, inline #390 of the F3 dataset. Note the correct delineation of the 7 types of horizons and the two stratigraphic sequences (in purple and white). The lower accuracy on the faults (in blue) and the salt dome (in yellow) can be improved by performing more training. Table 1 lists the corresponding confusion matrix for quantitative analysis.

when the training is terminated after 250 epochs.

After training the SpiNet, we then apply it to the 4 inline sections of manual annotations shown in Figure 2, including the three training sections (inline #190, 290, and 490) and the testing section (inline #390). Figure 6 and 7 show the corresponding prediction, respectively. Compared to the manual annotations, the 12 target seismic patterns are correctly identified, particularly these of geologic importance like the two stratigraphic sequences (in purple and white), the chaotic reflection (in magenta), the salt dome (in yellow) as well as the overlaying deformed reflection (in green). Table 2 lists the corresponding confusion matrix, in which the overall accuracy is estimated as 78%. The error mainly comes from the incomplete delineation of the saltbodies (Pattern No.10), which undesirably mixes the outer portions with the surrounding reflection (Pattern No.6) as observed in Figure 7. Further improvement is expected by performing more training and feeding more annotations on the "confusing" zones.

## APPLICATIONS

After training and testing the proposed SpiNet, in this section we provide two examples for illustrating its added values in assisting subsurface interpretation from 3D seismic data, which include (a) real-time annotation of a seismic volume and (b) SpiNet-based construction of more seismic interpretation networks.

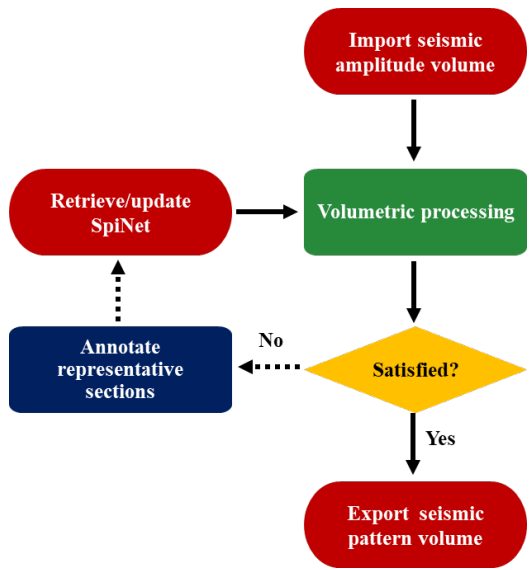

**Figure 8.** The workflow of utilizing the proposed SpiNet for seismic volume annotation. Such SpiNet-based volumetric annotation implements an iterative process for updating the SpiNet by manually annotating a few representative sections when the generated pattern volume is considered not satisfactory.

**Example 1: seismic volume annotation**

The SpiNet is built from the state-of-the-art deconvolutional neural network, and thus it successfully inherits its computational efficiency and is capable of quickly annotating a seismic volume into the 12 defined seismic patterns. Figure 8 displays the workflow for SpiNet-based seismic volume annotation, which consists of three steps. Specifically, given a seismic amplitude volume, first, the pre-trained SpiNet is retrieved and performed on the volume quickly generates the corresponding pattern volume. Next, the pattern volume is handed to an interpreter for result analysis. If the results are evaluated as less satisfactory, it is necessary for interpreters to annotate a few representative sections of the given seismic dataset and then update the SpiNet by feeding the newly-provided annotations. Such process is repeated until the prediction reaches high accuracy and become acceptable for interpreters. The capability of the SpiNet in end-to-end annotation ensures the applicability of the iterative process in practice. Finally, the pattern volume is exported for more advanced seismic interpretation and modeling.

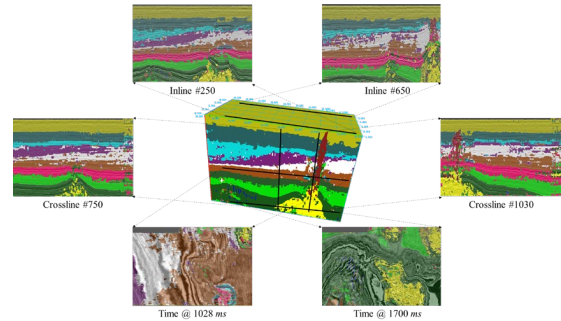

**Figure 9.** The 3D view of the pattern volume by applying the proposed SpiNet-based volumetric annotation workflow (Figure 8) to the entire F3 dataset. It is also clipped to 6 randomly-selected sections that were not seen at the stage of the SpiNet training. Note the correct labelling of the major features, such as salt domes (in yellow) as well as the overlaying deformed reflections (in magenta and green), the gas chimneys (in red), and the two stratigraphic sequences (in purple and white), all of which are considered of significant implications in this area.

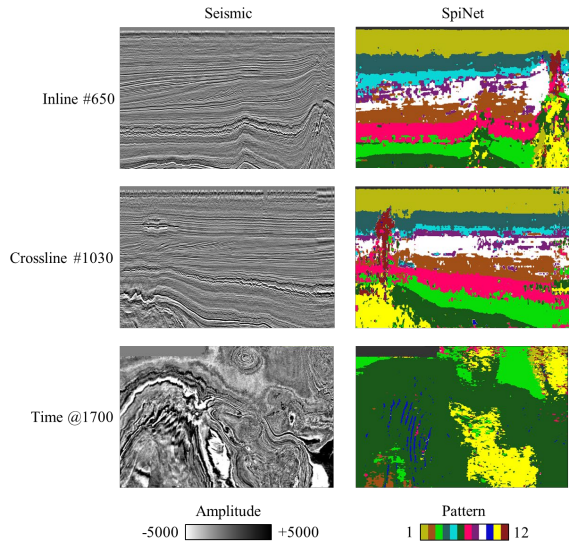

**Figure 10**. The prediction of 3 sections shown in Figure 9, including inline #650, crossline #1030, and time at 1700 ms. Note the correct labelling of the major features, such as salt domes (in yellow) as well as the overlaying deformed reflections (in magenta and green), the faults (in blue), and the gas chimney (in red).

Take the F3 seismic volume for example, which contains 651 inlines, 951 crosslines, and 463 sample per trace. Figure 9 displays a 3D view of the generated pattern volume as well as its clipping to 6 randomly-selected sections, none of which has been seen by the SpiNet at the training stage. For the convenience of better quality control, we then display the prediction of 3 sections, including inline #650, crossline #1030, and time slice at 1700 *ms* in Figure 10. Apparently, the seismic patterns are correctly interpreted in

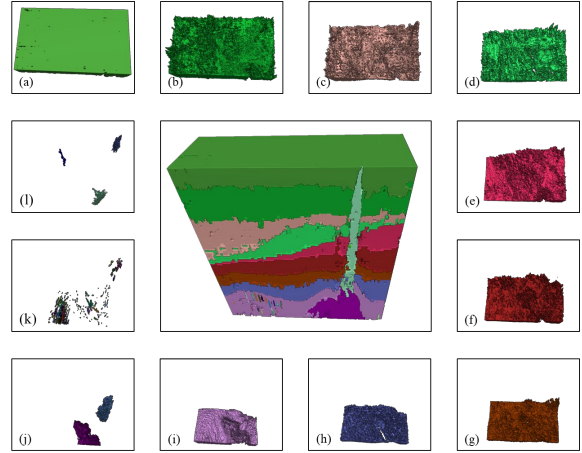

**Figure 11**. An example of utilizing the pattern volume generated by the proposed SpiNet-based seismic volume annotation workflow (Figure 8), which successfully separates each of these patterns as an individual geobody and builds the corresponding geologic model.

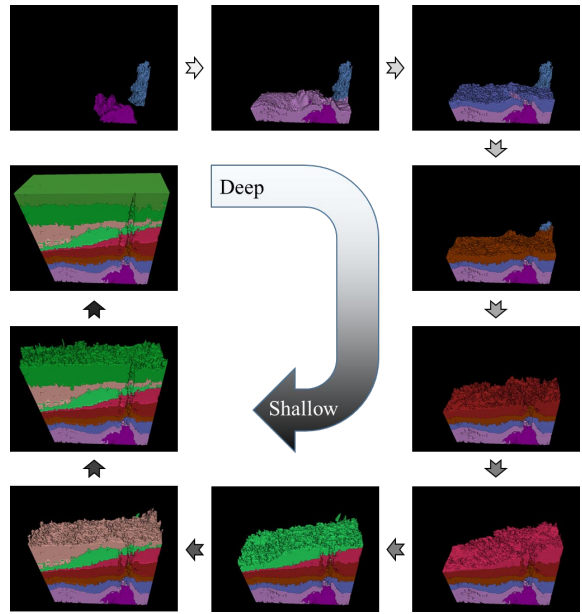

**Figure 12**. The modeling of the deposition in the F3 block based on the the proposed SpiNet-based seismic volume annotation (Figure 8).

general, particularly these important structures like the salt domes (in yellow), the faults (in cyan), the gas chimney (in maroon),

the chaotic reflection (in red), and the two stratigraphic sequences (in purple and white) that are considered of important depositional implications in this area. Such automated volume annotation is consistent with previous studies in this area, particularly the salt dome and faults delineated in the time section. Integrated with the powerful seismic interpretation tools such as geobody extraction, the pattern volume (Figure 9) allows interpreters to readily separate each depositional event (Figure 11) and restore the geologic history in the F3 area (Figure 12). In term of the computational efficiency, it takes about 10 minutes for annotating the entire F3 volume by using one piece of NVidia P4000 GPU.

**Example 2: Fault detection**

For achieving the goal of seismic pattern identification and annotation, the proposed SpiNet internally generates a series of features from its convolutional and deconvolutional layers, each of which understands the input seismic signals in its unique way. Although seismic signals vary from one area to another in detail, the fundamentals are the same, and thus the seismic patterns are similar in general. For example, a saltbody is featured with chaotic reflection in it, and faults are featured with apparent offset that breaks lateral continuity of seismic reflection. Therefore, the capability of the SpiNet in understanding seismic signals is not limited to the training dataset SpiNet only but can be applied to any seismic volume collected from various areas, which indicates the feasibility of utilizing the trained SpiNet to build new seismic interpretation networks that are more task-oriented and even identify more seismic patterns not defined in Figure 1 or beyond the scope of the current SpiNet.

Take fault detection for example. Although fault is one of the 12 seismic patterns (Figure 1), the geologic complexities in the subsurface make it still challenging for identifying all types of faults, particularly if the faults in the target dataset have not been covered in the SpiDat and seen by the SpiNet. Therefore, directly performing the SpiNet would generate the results that are less satisfactory in accuracy and with many false positives as shown in Figure 14a. In such a case, a new fault detection network (FaultNet) is in need for completing the specific interpretation task. Figure 13 displays the workflow for SpiNet-based FaultNet construction, which consists of five steps. Specifically, given a seismic amplitude volume, first, the FaultNet is designed based on the pre-trained SpiNet. Then, a few representative sections in the seismic volume are selected and manually annotated, and the FaultNet is trained by feeding the manual annotations. Third, performing the trained FaultNet on the seismic volume quickly generates the corresponding fault volume.

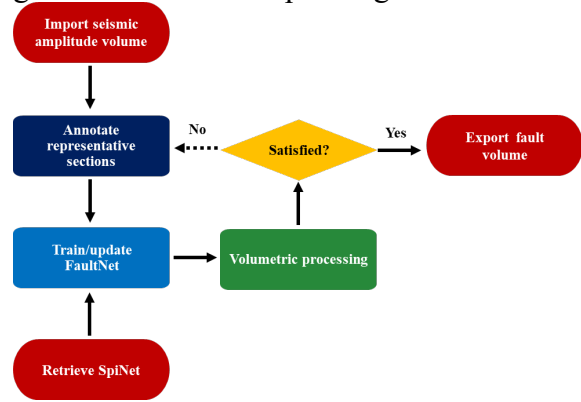

**Figure 13**. The workflow of utilizing the proposed SpiNet for building a fault detection network (FaultNet). The major superiority of such SpiNet-based fault detection comes from the pass of the SpiNet capability of understanding seismic signals to the FaultNet, so that the latter could be trained more efficiently. A comparison between training the FaultNet from scratch and the SpiNet is shown in Figure 16.

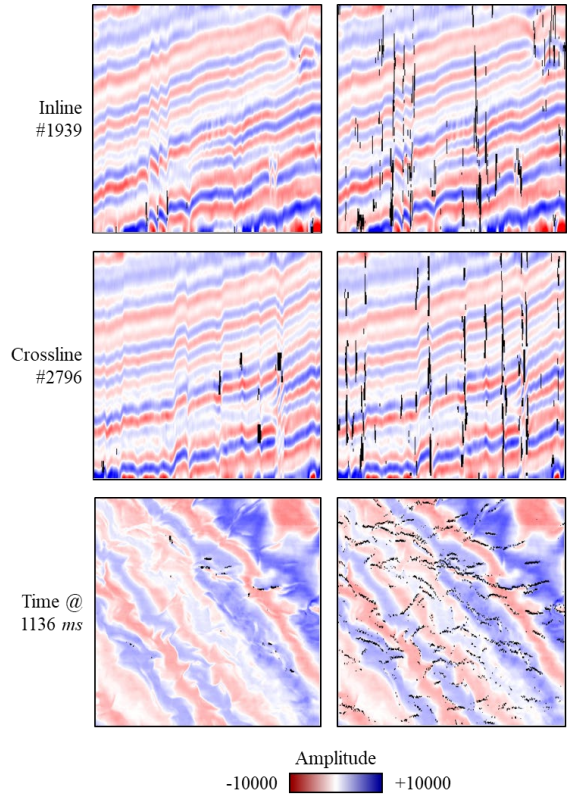

**Figure 14**. The improved accuracy of fault detection by applying the proposed SpiNet (a) and the SpiNet-based FaultNet (b) to the 3D seismic dataset from Great South Basin (GSB) in New Zealand. The former provides only partial detection, since the target faults have not been seen by the SpiNet, whereas the SpiNet-based FaultNet leads to significantly improved detection accuracy, since the target fault patterns are feed into the FaultNet training (Figure 13).

Next, the fault volume is reviewed by interpreters for result analysis. If the fault detection is considered less satisfactory, it is recommended for interpreters to annotate a few representative sections and then update the FaultNet by feeding these new annotations. Such process is repeated until the detected faults achieve high accuracy. Finally, the fault volume is exported for assisting more advanced fault interpretation and structural modeling.

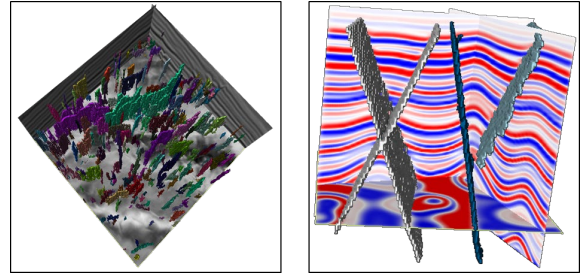

**Figure 15**. The fault interpretation by applying the generated SpiNet-based FaultNet to the GSB dataset (a) and a new dataset (b).

For demonstrating the efficiency of the proposed workflow, we use a subset of 3D seismic dataset from the Great South Basin in New Zealand, which contains 483 inlines, 501 crosslines, and 76 sample per trace and is featured with polygonal faults. Figure 14b displays the fault lineaments detected by the SpiNet-based FaultNet, which has a significantly improved accuracy compared to the detection simply from the SpiNet. With its performance validated, the FaultNet is not only capable of detecting the polygonal faults in the GSB area (Figure 15a), but more importantly can then be used for fault interpretation from other new datasets by integrating the transfer learning if necessary (Figure 15b).

Compared to the building of a seismic interpretation network from scratch, the major superiority of the SpiNet-based workflow comes from the inherent capability of understanding seismic signals inherited from the SpiNet, which contributes to faster convergence at the stage of network training. Such superiority is illustrated in Figure 16, which compares the loss curve of training the FaultNet from scratch (in blue) and the SpiNet (in red) in 100 epochs. Apparently, the initial loss is high when the FaultNet is trained from scratch, which is reasonable as the network knows nothing about the seismic signals and labels and its predictions are majorly from guess. On the contrary, the initial loss of the SpiNet-based FaultNet is much lower, and correspondingly the entire training process is more efficient.

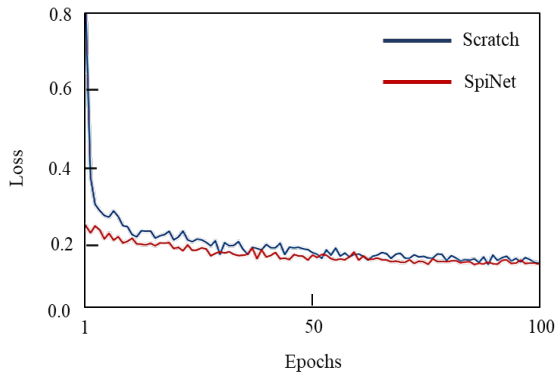

**Figure 16**. The comparison of the loss curves for training the fault detection network (FaultNet) from scratch (in blue) and the proposed SpiNet (in red). Apparently, the SpiNet-based FaultNet training starts with a significantly reduced loss and converges faster, which results from the successful pass of the capability of understanding seismic signals from the SpiNet to the FaultNet.

## CONCLUSIONS

We have developed a seismic pattern interpretation network (SpiNet) for assisting 3D seismic interpretation. Compared to the existing tools, it is superior in three aspects. First, it is capable of identifying 12 common seismic patterns, including these of geologic importance like saltbodies, faults, and stratigraphic sequences. Second, it directly works on a seismic section and thereby is capable of annotating a seismic volume at real time. Third, it is applicable to any seismic datasets of any dimensions without spending much time and efforts on data preconditioning. Besides its efficiency of volumetric seismic annotation, the proposed SpiNet also paves the foundation for developing new seismic interpretation networks that are more task-oriented and cover more seismic patterns.

Together with the SpiNet, we also build a seismic pattern interpretation dataset (SpiDat) for further work. However, due to our limited dataset resources and geoscientific knowledge, our annotation may be not comprehensive enough for coving all important seismic features or undesirably mixture a few features into one pattern. More work is no doubt in need for further improving the list of seismic patterns as well as their corresponding annotations. Besides the real seismic datasets, generating synthetic ones is also complementary for expanding SpiDat, which could be model-based (e.g., Wu and Hale, 2016) or by the generative adversarial networks (Goodfellow, 2014).

The current architecture of the SpiNet consists of a total of 12 layers (Figure 3) and targets 12 seismic patterns (Figure 1), which may be considered not efficient and deep enough particularly when the training dataset SpiDat is increasing and/or more seismic patterns are added into the target list in the future. Therefore, it is highly anticipated that the SpiNet would be re-designed and trained, in which the current version can be used as the basis for the SpiNet upgrading.

## ACKNOWLEDGMENTS

This work is supported by the Center for Energy & Geo Processing (CeGP) at Georgia Institute of Technology and King Fahd University of Petroleum and Minerals. We would like to thank Open Seismic Repository, the New Zealand Petroleum and Minerals (NZP&M), and Dr. Xinming Wu for providing the F3 block over the Netherlands North Sea, the Great South Basin (GSB) dataset in New Zealand, and the synthetic fault volume, respectively. The neural network algorithm is implemented based on the open-source Python package *TensorFlow* developed by Google Brain.

**Table 1.** The tentative categorization of seismic patterns used for building the seismic pattern interpretation dataset (SpiDat) to facilitate the implementation of machine learning into the domain of seismic interpretation. Note that the unpredictable geologic complexities and varying interpretational goals add the difficulty of cohering the descriptors, and correspondingly, the pattern list cannot be easily finalized.

| Category | Descriptor | Seismic pattern | Example |
|---|---|---|---|
| Horizon | a. geometry (e.g., chaotic, dipping) <br> b. intensity (e.g., strong) <br> c. …… | a. chaotic <br> b. 15°-dipping <br> c. flat strong <br> d. …… | 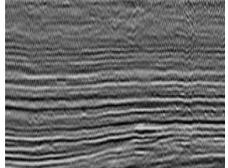 |
| Structure | a. deformation (e.g., faulting) <br> b. scale (e.g., large) <br> c. …… | a. fault <br> b. saltbody <br> c. anticline <br> d. …… | 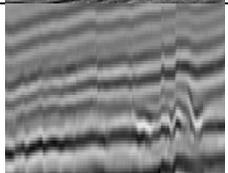 |
| Stratigraphy | a. morphology (e.g., transgressive) <br> b. intensity (e.g., strong) <br> c. …… | a. transgression <br> b. regression <br> c. disconformity <br> d. …… | 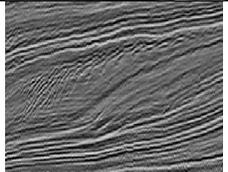 |

**Table 2.** The confusion matrix of applying the trained SpiNet to the testing section (Inline #390) between the prediction (P) and the manual annotations (A) for the 12 defined seismic patterns in Figure 1. The overall pixel-wise annotation accuracy is 78%.

|  | P-1 | P-2 | P-3 | P-4 | P-5 | P-6 | P-7 | P-8 | P-9 | P-10 | P-11 | P-12 |
|---|---|---|---|---|---|---|---|---|---|---|---|---|
| A-1 | **63140** | 45 | 239 | 5268 | 139 | 1096 | 0 | 176 | 39 | 15 | 140 | 40 |
| A-2 | 0 | **30896** | 1007 | 24 | 170 | 528 | 448 | 1571 | 2561 | 0 | 0 | 0 |
| A-3 | 0 | 572 | **34478** | 0 | 10 | 2299 | 5107 | 114 | 38 | 67 | 431 | 490 |
| A-4 | 2216 | 2 | 0 | **38944** | 1701 | 0 | 18 | 259 | 42 | 0 | 0 | 138 |
| A-5 | 5 | 339 | 0 | 9509 | **19345** | 0 | 0 | 1994 | 360 | 27 | 0 | 34 |
| A-6 | 43 | 880 | 1927 | 32 | 1 | **50960** | 1188 | 0 | 164 | 834 | 1515 | 317 |
| A-7 | 11 | 3395 | 2403 | 80 | 53 | 1813 | **35011** | 42 | 255 | 209 | 186 | 103 |
| A-8 | 0 | 1515 | 98 | 1356 | 5353 | 44 | 124 | **22771** | 3749 | 19 | 0 | 0 |
| A-9 | 0 | 6299 | 196 | 57 | 1286 | 311 | 108 | 7821 | **28743** | 107 | 0 | 0 |
| A-10 | 0 | 0 | 43 | 0 | 0 | 1304 | 109 | 0 | 0 | 648 | 0 | 0 |
| A-11 | 57 | 0 | 2424 | 0 | 0 | 7353 | 384 | 0 | 0 | 426 | **8484** | 751 |
| A-12 | 0 | 0 | 0 | 0 | 0 | 0 | 0 | 0 | 0 | 0 | 0 | 0 |